\documentclass[10pt,conference,letterpaper]{IEEEtran}
\IEEEoverridecommandlockouts

\usepackage{cite}
\usepackage{amsmath,amssymb,amsfonts}
\usepackage{algorithmic}
\usepackage{graphicx}
\usepackage{textcomp}
\usepackage{xcolor}
\usepackage{booktabs}
\usepackage{multirow}
\usepackage{graphicx}
\usepackage{colortbl}
\usepackage{makecell}

\def\BibTeX{{\rm B\kern-.05em{\sc i\kern-.025em b}\kern-.08em
    T\kern-.1667em\lower.7ex\hbox{E}\kern-.125emX}}
\begin{document}

\title{Deep Semantic Inference over the Air: An Efficient Task-Oriented Communication System\\

\thanks{The work was supported in part by the Horizon Europe MSCA Staff Exchanges Project, IPOSEE (Intelligent and Proactive Optimisation for Service-centric Wireless Networks), grant agreement ID: 101086219. The data handling was enabled by resources provided by the National Academic Infrastructure for Supercomputing in Sweden (NAISS), partially funded by the Swedish Research Council through grant agreement no. 2022-06725.}
}

\author{\IEEEauthorblockN{Chenyang Wang, Roger Olsson, Stefan Forsström, Qing He}

\IEEEauthorblockA{Department of Computer and Electrical Engineering, Faculty of Science, Technology and Media, \\ Mid Sweden University, Sundsvall, Sweden \\ Email: \textit{\{chenyang.wang, roger.olsson, stefan.forsstrom, qing.he\}@miun.se}}}

\maketitle

\begin{abstract}

Empowered by deep learning, semantic communication marks a paradigm shift from transmitting raw data to conveying task-relevant meaning, enabling more efficient and intelligent wireless systems. In this study, we explore a deep learning-based task-oriented communication framework that jointly considers classification performance, computational latency, and communication cost. We evaluate ResNets-based models on the CIFAR-10 and CIFAR-100 datasets to simulate real-world classification tasks in wireless environments. We partition the model at various points to simulate split inference across a wireless channel. By varying the split location and the size of the transmitted semantic feature vector, we systematically analyze the trade-offs between task accuracy and resource efficiency. Experimental results show that, with appropriate model partitioning and semantic feature compression, the system can retain over 85\% of baseline accuracy while significantly reducing both computational load and communication overhead.

\end{abstract}


\section{Introduction}

Traditional wireless communication systems have been designed according to Shannon’s information theory, where the primary objective is to reliably transmit bits over noisy channels \cite{wcom}. However, with the rapid rise of data-driven applications such as autonomous driving, real-time surveillance, and industrial automation, communication networks are now expected to support not only reliable data transfer but also intelligent, task-specific decision-making. This paradigm shift has prompted a rethinking of communication system design, leading to growing interest in semantic and task-oriented communication \cite{10554663}, which focuses on conveying information relevant to the underlying task, aligning more closely with how humans communicate efficiently and robustly in uncertain environments \cite{6773024,9919752,10855638}.

Recent advances in machine learning, particularly deep learning, have significantly accelerated the development of the semantic communication systems. Deep neural networks (DNNs), such as convolutional neural networks (CNNs) for images or Transformers for language, can capture high-level abstractions. Semantic encoders and decoders can be co-trained with downstream inference tasks to extract and transmit only the most relevant features. For example, instead of sending the full-resolution image, the encoder may learn to transmit only task-critical semantic features (e.g., class labels, bounding boxes, or embeddings), greatly reducing communication cost without compromising task performance. This concept extends naturally to task-oriented communication \cite{9955525}, where performance is measured not by bit error rate but by how well a specific task (e.g., classification, detection, or control) is completed. Such systems are especially valuable in resource-constrained environments like wireless sensor networks or mobile edge computing, where devices have limited energy, bandwidth, and processing power.

Despite recent progress, designing and deploying deep learning-based task-oriented communication systems remains challenging. These systems must integrate sensing, encoding, transmission, and inference into a unified pipeline, where model architectures and training objectives must align with system-level constraints such as latency, energy consumption, and bandwidth. While deep models can be trained offline using powerful infrastructure, real-time inference during deployment must be both accurate and efficient, requiring a careful balance between task performance and the use of computational and communication resources. Existing studies address only parts of this challenge \cite{10.1145/3037697.3037698, 9145068, 10226176, 10287247}. For instance, \cite{10.1145/3037697.3037698} introduced adaptive device–edge co-inference to better utilize distributed computing resources but did not consider communication cost and channel conditions. Later work in \cite{9145068} focused on compressing intermediate features to reduce transmission load, yet treated model partitioning as a secondary concern. More recent research has examined the timeliness of task-oriented inference via Age of Information \cite{10226176}, but without investigating the optimal split decisions. Overall, a unified analysis of supervised learning models under different partitioning settings considering both channel noise and compression network complexity remains lacking.

In this paper, we present a deep learning-based task-oriented communication framework that explicitly considers both task accuracy and latency. The latter includes computation delay (from model inference) and communication delay (from transmitting semantic features). We focus on classification tasks, adopting the 18-layer and 34-layer Residual Networks \cite{7780459} as backbone models and evaluating performance on the CIFAR-10 and CIFAR-100 datasets \cite{krizhevsky2009learning}. Through extensive experiments, we investigate how different model partitioning strategies and semantic feature dimensions influence the trade-offs among accuracy, computational load, and communication cost, providing practical insights for the design of efficient task-oriented communication systems.

The remainder of this paper is organized as follows. Sec.~II introduces the system architecture. Sec.~III presents the design and implementation of the classification system. Sec.~IV evaluates the system under various configurations and analyzes the trade-offs between accuracy and latency. Sec.~V concludes the paper and discusses future directions.

\section{Task Latency in Deep Learning-Based Task-Oriented Communication}

We consider a deep learning-based task-oriented communication system designed to perform classification tasks. Depending on how the DNN model is deployed, three typical configurations can be considered:

\begin{itemize}
    \item Transmitter-side inference: The classifier is fully deployed at the transmitter (edge device), and only the final classification result (label) is sent. This minimizes communication overhead but places high computational demands on resource-constrained edge devices. 
    \item Receiver-side inference: The full input is transmitted to the receiver (cloud device), which performs the classification. This reduces transmitter-side computation but incurs high communication cost.
    \item Split inference: The DNN model is partitioned between the transmitter and the receiver, allowing task-relevant semantic features to be extracted at the edge and further processed in the cloud. This approach balances communication and computational efficiency.
\end{itemize}

\begin{figure}
    \centering
    \includegraphics[width=1\linewidth]{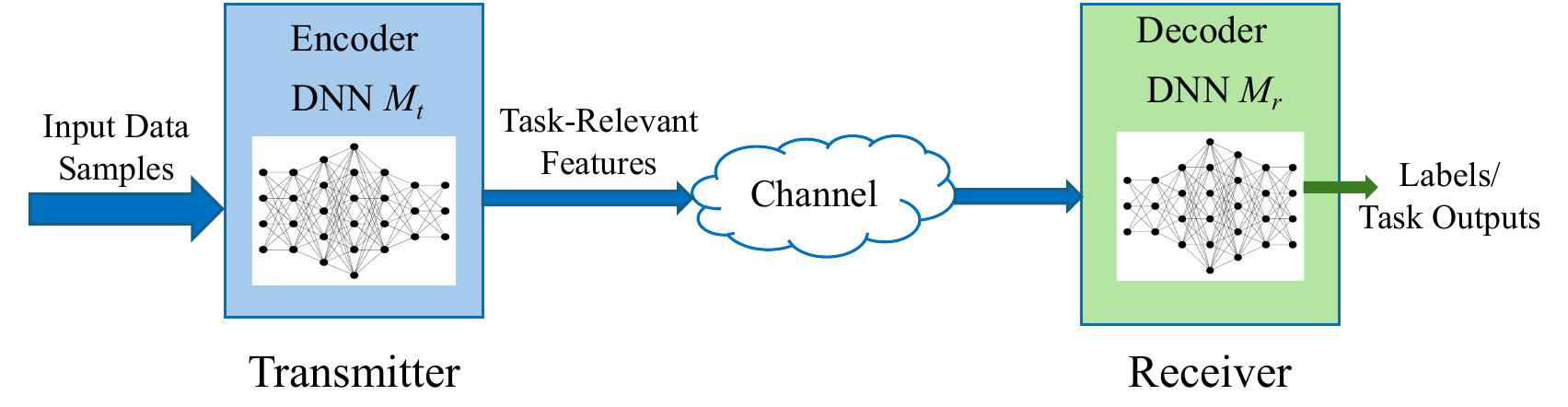}
    \vspace{-4mm}
    \caption{A deep learning-based task-oriented communication network.}
    \label{fig:arch}
\end{figure}

We illustrate the system architecture in Fig.~\ref{fig:arch}. Let $M_t$ and $M_r$ denote the parts of the DNN model deployed at the transmitter and receiver, respectively. Their inference times are denoted as $T_{M_t}$ and $T_{M_r}$. In general, the inference time of a model depends on both its complexity and the hardware it runs on. We quantify model complexity using floating-point operations (FLOPs), denoted by $F_M$ for the model $M$, which provides a hardware-agnostic measure of computational demand. The actual computation time of the model $M$ can be estimated as a function of its FLOP counts, where the function is determined by the hardware settings of the device that runs the model. Inside the compute-bound regime, the function is linear and monotone, so we estimate the computation time as:
\begin{equation}
\label{eq: tm}
    T_{M_t} = \alpha_t F_{M_t}, \quad
    T_{M_r} = \alpha_r F_{M_r},
\end{equation}
where $\alpha_t$ and $\alpha_r$ are device-specific coefficients representing the per-FLOP processing time at the transmitter and receiver, respectively. The total computation time is therefore:
\begin{equation}
\label{eq: tcomp}
    T_{\text{comp}} = T_{M_t} + T_{M_r} = \alpha_t F_{M_t} + \alpha_r F_{M_r}.
\end{equation}

For the communication part, let $N_c$ denote the size (in bits or bytes) of the intermediate semantic feature representation produced by $M_t$, and let $R$ be the transmission rate of the communication channel. Then the communication time is:
\begin{equation}
\label{eq:tcomm}
    T_{\text{comm}} = \frac{N_c}{R}.
\end{equation}

Combining both components, we define the overall inference time for a task execution as:
\begin{equation}
\label{eq:ttask}
    T_{\text{task}}  = T_{\text{comp}} + T_{\text{comm}}.
\end{equation}

Based on the equations $\eqref{eq: tm}$ - $\eqref{eq:ttask}$, if the devices are fixed and communication settings, such as bandwidth and modulation scheme, are predetermined, then the selection of the model partitioning strategy and the size of the intermediate semantic representation $N_c$ jointly determine whether the task can be executed accurately and timely.

In the following sections, we design a task-oriented communication system for classification, and evaluate the trade-off between latency and accuracy under different model partitioning strategies and $N_c$ values.

\section{System Design for Task-Oriented Classification}

\begin{figure*}[htbp]
\begin{center}
\includegraphics[width=0.88\linewidth]{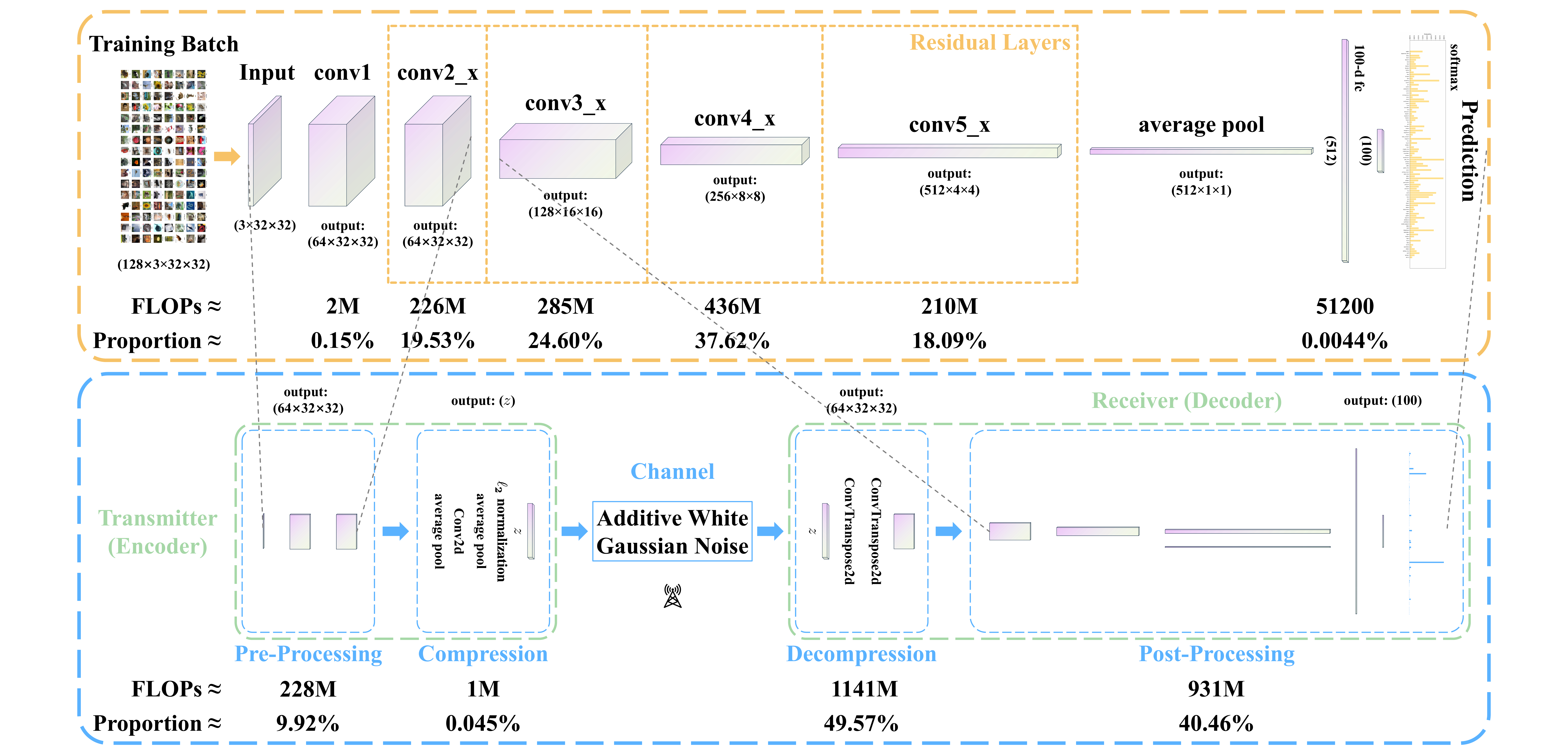}
\end{center}
\vspace{-4mm}
\caption{Illustration of ResNet-34 architecture and a split-inference task-oriented network based on the ResNet-34 Vanilla model (unmodified baseline version). Specifically, the first three parts (input, conv1, and conv2\_x) are considered in the transmitter (encoder) as pre-processing, while the rest of them are considered in the receiver (decoder) as post-processing. The compression and decompression modules ensure the post-processing input is consistent with the pre-processing output while preserving invariance and equivariance to translation. The semantic representation with $z$ dimension is transmitted from the encoder to the decoder over the AWGN channel. The FLOPs and proportion of each module are listed below. (Zoomed in for better visualization.)}
\label{fig:resnet}
\end{figure*}

\subsection{Deep Learning Model for Classification}

We adopt the 18-layer and the 34-layer Residual Networks (ResNet-18 and ResNet-34) as the backbone models for our task-oriented communication system. These widely used CNN architectures are known for their robustness, modularity, and strong performance across a broad range of image classification tasks.

For CIFAR-100, as shown in Fig.~\ref{fig:resnet}, we use customized variants of ResNet-18 and ResNet-34, adapted to the $32 \times 32$ image resolution. The original $7 \times 7$ convolutional and max pooling layers are replaced with a single $3 \times 3$ convolutional layer (stride 1, padding 1) with 64 filters, preserving spatial resolution. The output layer of the fully connected classifier is modified to predict 100 classes instead of the original 1000.
 
These networks follow a hierarchical structure: shallow layers extract low-level features such as edges and textures, while deeper layers capture higher-level semantic representations. ResNet-34, in particular, offers a good balance between classification accuracy and model complexity, making it suitable for edge–cloud deployment. While the standard ResNet-34 (for $224 \times 224$ images) contains approximately 21.8 million parameters and 3.6 billion FLOPs, the CIFAR-100 variant is significantly lighter, with about 21.3 million parameters and 1.16 billion FLOPs.

Due to the widespread adoption and adaptability across datasets and domains, the ResNet-based architectures like ResNet-34 make our system design generalizable to a variety of classification tasks and application scenarios.

\subsection{Model Partitioning}
To integrate ResNet-34 into a task-oriented communication system, we partition the model into two components: an encoder $M_t$ at the transmitter (e.g., an edge device) and a decoder $M_r$ at the receiver (e.g., a cloud server or base station), as shown in Fig.~\ref{fig:resnet}. The encoder’s output, $z \in \mathbb{R}^{N_c}$, is an intermediate semantic representation transmitted to the decoder $M_r$ over the communication channel. Here, $N_c$ refers to the dimensionality of $z$, which corresponds to the number of neurons in the last layer of $M_t$ (and the input of $M_r$). This dimension determines the amount of semantic information transferred and thus plays a critical role in balancing task performance against communication overhead.

\subsection{Wireless Channel Simulation with AWGN}
To simulate the effect of a wireless communication channel, we insert an Additive White Gaussian Noise (AWGN) layer between $M_t$ and $M_r$. This layer introduces stochastic perturbations to the semantic features after $\ell_2$ normalization and scaling, emulating the degradation commonly observed in wireless transmission. Formally, the transmitted semantic feature vector $z \in \mathbb{R}^{N_c}$ is corrupted by Gaussian noise:
\begin{equation}
    \label{eq:noise}
    \tilde{z} = z + n, \quad n \sim \mathcal{N}(0, \sigma^2 I),
\end{equation}
where $\sigma^2$ denotes the noise variance and is controlled by the assigned signal-to-noise ratio (SNR) of the channel. The receiver obtains the noisy feature vector $\tilde{z}$ and restores it to its pre-compression size, then uses it as the input to $M_r$ to perform the remaining convolutional layers and finally the classification task. This setup enables us to assess the robustness and performance of the system under varying channel conditions. To ensure a seamless interface between the encoder and decoder, the output dimension of $M_t$ should match the input dimension of $M_r$, i.e., both must operate on a vector of size $N_c$. Training and evaluation are implemented in an end-to-end manner within the widely adopted deep learning framework, PyTorch \cite{pytorch}. \textit{The source code is available at the GitHub repository: https://github.com/sanyeungwang/Deep-Semantic-Inference-over-the-Air}.

\section{Performance Evaluation}

\subsection{Datasets}

We evaluate our system using two widely adopted image classification benchmarks: CIFAR-10 and CIFAR-100.

\textit{CIFAR-10:} This dataset \cite{krizhevsky2009learning} contains 60,000 color images of size $32 \times 32$ across 10 classes: airplane, automobile, bird, cat, deer, dog, frog, horse, ship, and truck. Each class includes 6,000 images. The dataset is divided into five training batches and one test batch, each with 10,000 images. Each RGB image has a shape of $32 \times 32 \times 3$, and class labels are encoded as integers from 0 to 9.

\textit{CIFAR-100:} Also introduced in \cite{krizhevsky2009learning}, CIFAR-100 shares the same image format as CIFAR-10 but includes 100 classes with 600 images each. For every class, there are 500 training images and 100 test images, making the classification task more fine-grained. The dataset is split into a single training batch of 50,000 images and a test batch of 10,000 images.

\subsection{Implementation Details}
Our experiments are conducted on both CIFAR-10 and CIFAR-100 datasets using PyTorch. Data augmentation is performed through random cropping with a padding of 4 and random horizontal flipping, followed by normalization, as specified by the Transforms. All models are trained with a batch size of 128, a total of 100 epochs, and an initial learning rate of 0.1. Stochastic Gradient Descent (SGD) with a momentum of 0.9 and a weight decay of 5e-4 is leveraged for optimization. The cosine annealing learning rate scheduler is employed to gradually reduce the learning rate to zero. All experiments are conducted using an NVIDIA RTX 3090 GPU. Reproducibility is ensured by deterministic algorithms and fixed random seeds. The reported Top-1 Accuracy is obtained from the best validation checkpoint during training. The standard deviation $\sigma$ of the AWGN, controlled by the assigned SNR in decibels, is calculated as:
$\sigma = \frac{1}{\sqrt{10^{\text{SNR}_\text{dB} / 10}}}.$

\subsection{Comparisons With Different Settings}

\begin{figure}[!ht] 
	\begin{center}
		\includegraphics[width=1\linewidth]{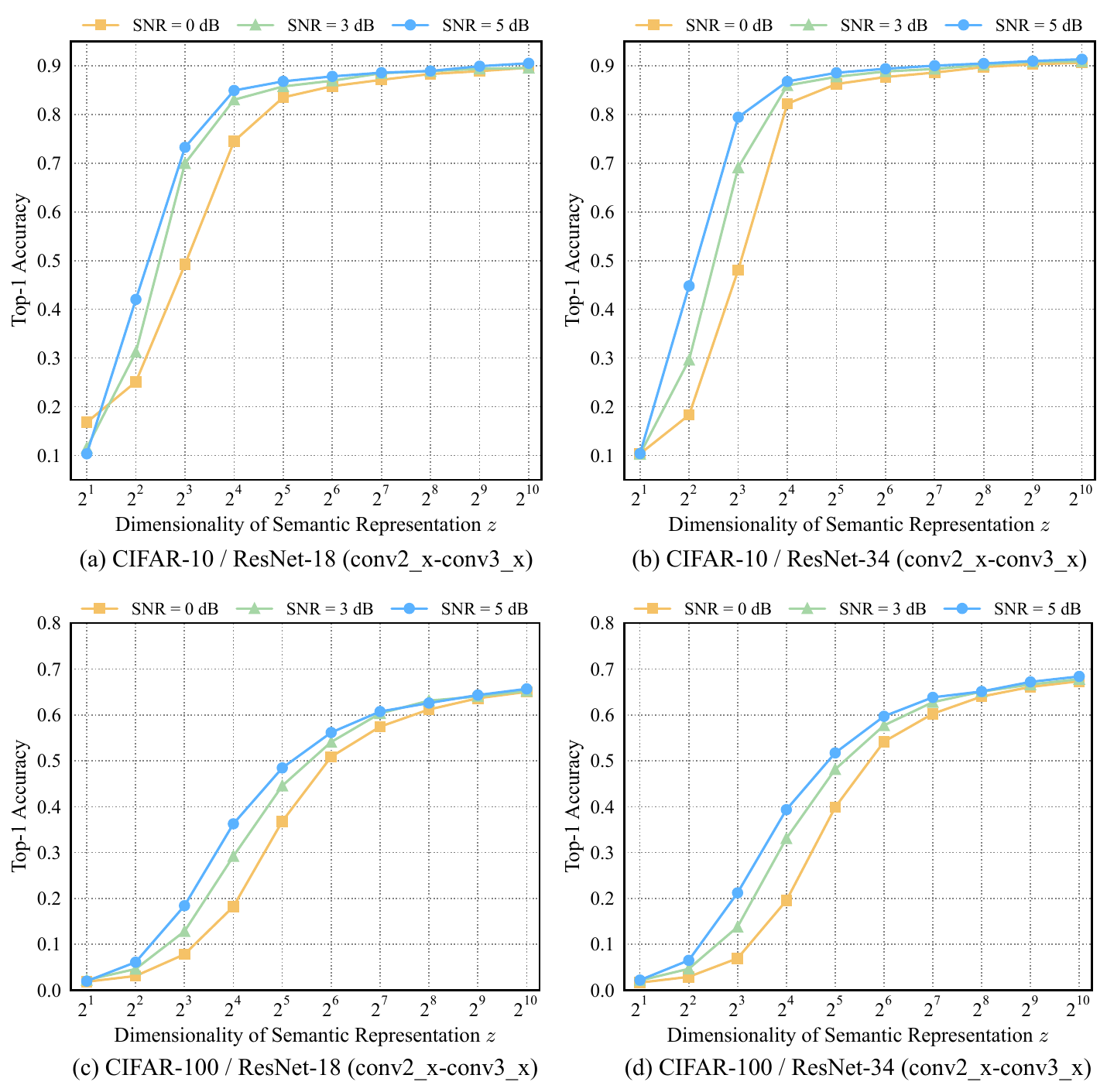}
	\end{center}
    \vspace{-4mm}
	\caption{Comparison of Top-1 Accuracy on CIFAR-10 and CIFAR-100 using ResNet-18 (Base + Split (SP-2) + AWGN) and ResNet-34 (Base + Split (SP-2) + AWGN) across different $z$ dimensions (implemented by varying the number of channels $N_c$) and SNR levels. Note that the names of the models above and the results on CIFAR-100 for SNR = 5 dB and $N_c=1024$ are consistent with those in Table \ref{vanilla}.}
        \label{fig:cifar10}
\end{figure}

Based on the proposed task-oriented communication system, we first analyze performance under different $z$ dimensions and SNR levels, as shown in Fig.~\ref{fig:cifar10}. We evaluate both ResNet-18 and ResNet-34, each split between \textit{conv\_2} and \textit{conv\_3} (referred to as SP-2), with one fractionally-strided convolution module (ConvTranspose2d) in the decompression module, and assuming unlimited computing power in the cloud device initially. Top-1 Accuracy refers to the comparison between the class with the highest probability and the ground-truth label. For CIFAR-10 shown in Figs.~\ref{fig:cifar10}(a) and \ref{fig:cifar10}(b), we observe that higher SNR leads to better performance, but accuracy approaches 0.9 even for low-dimensional $z$. This saturation effect makes CIFAR-10 less suitable for analyzing the impact of model architecture and compression settings. In contrast, CIFAR-100 (Figs.~\ref{fig:cifar10}(c) and \ref{fig:cifar10}(d)) achieves lower overall accuracy due to its finer-grained classes and smaller inter-class variance. The performance gap between different SNR levels is also more pronounced, highlighting its sensitivity to AWGN and its suitability for studying the effects of semantic compression and noise. Therefore, we focus on CIFAR-100 in subsequent experiments.

To further compare the effects of different settings and model structures, we expand our experiments to obtain the results summarized in Table~\ref{vanilla}, with $SNR$ and $N_c$ uniformly set to 5 dB and 1024, respectively. Specifically, the vanilla model (unmodified baseline) with transforms is regarded as the base model. Experimental results for several split points (as defined in Table~\ref{vanilla}), with and without AWGN, are reported. Compared with ResNet-18, ResNet-34 achieves higher accuracy and, as a deeper model, better demonstrates the benefits of computational separation between transmitter and receiver. For example, when both models are split at SP-2, the accuracy degradation caused by AWGN is smaller for ResNet-34 (from 0.7789 to 0.6838) than for ResNet-18 (from 0.7733 to 0.6567). Based on these observations, we focus more on ResNet-34 in the subsequent experiments to analyze the impact of different split locations.

\begin{table}[!ht]
\caption{Comparison of Top-1 Accuracy on CIFAR-100 under different settings, given SNR = 5 dB and $N_c$ = 1024. Gray shading denotes the same split point in Fig.~\ref{fig:resnet}. $^\dag$ indicates the identical models or models identical except for the split point.}
\vspace{-2mm}
\label{vanilla}
\centering
\begin{tabular}{c|l|c}
\toprule
\textbf{Model} & \multicolumn{1}{c|}{\textbf{Setting}} & \textbf{Top-1 Accuracy} \\ \midrule\midrule
\multirow{6.5}{*}{ResNet-18} & Vanilla & 0.6313 \\
& \cellcolor{gray!30}Vanilla + Split (SP-2) & \cellcolor{gray!30}0.6589 \\
& \cellcolor{gray!30}Vanilla + Split (SP-2) + AWGN & \cellcolor{gray!30}0.6428 \\ \cmidrule{2-3}
& Vanilla + Transforms (Base) & \textbf{0.7733} \\
& \cellcolor{gray!30}Base + Split (SP-2) & \cellcolor{gray!30}0.6801 \\
& \cellcolor{gray!30}Base + Split (SP-2) + AWGN & \cellcolor{gray!30}0.6567 \\ \midrule
\multirow{6.5}{*}{ResNet-34} & Vanilla & 0.6476 \\
& \cellcolor{gray!30}Vanilla + Split (SP-2) & \cellcolor{gray!30}0.6750 \\
& \cellcolor{gray!30}Vanilla + Split (SP-2) + AWGN & \cellcolor{gray!30}0.6705 \\ \cmidrule{2-3}
& Vanilla + Transforms (Base) & \textbf{0.7789} \\
& \cellcolor{gray!30}Base + Split (SP-2) & \cellcolor{gray!30}0.6952 \\
& \cellcolor{gray!30}Base + Split (SP-2) + AWGN $^\dag$ & \cellcolor{gray!30}0.6838 \\ \midrule
\multirow{7}{*}{Split Point} & SP-0 (train\_loader-conv1) $^\dag$ & 0.5600 \\
& SP-1 (conv1-conv2\_x) $^\dag$ & 0.4298 \\
& \cellcolor{gray!30}SP-2 (conv2\_x-conv3\_x) $^\dag$ & \cellcolor{gray!30}0.6838 \\
& SP-3 (conv3\_x-conv4\_x) $^\dag$ & 0.7560 \\
& SP-4 (conv4\_x-conv5\_x) $^\dag$ & \textbf{0.7753} \\
& SP-5 (conv5\_x-avgpool, fc) $^\dag$ & 0.7717 \\
& SP-6 (output\_logits-argmax) $^\dag$ & 0.6191 \\ \bottomrule
\end{tabular}
\end{table}

\begin{figure*}[htbp]
	\begin{center}
		\includegraphics[width=1\linewidth]{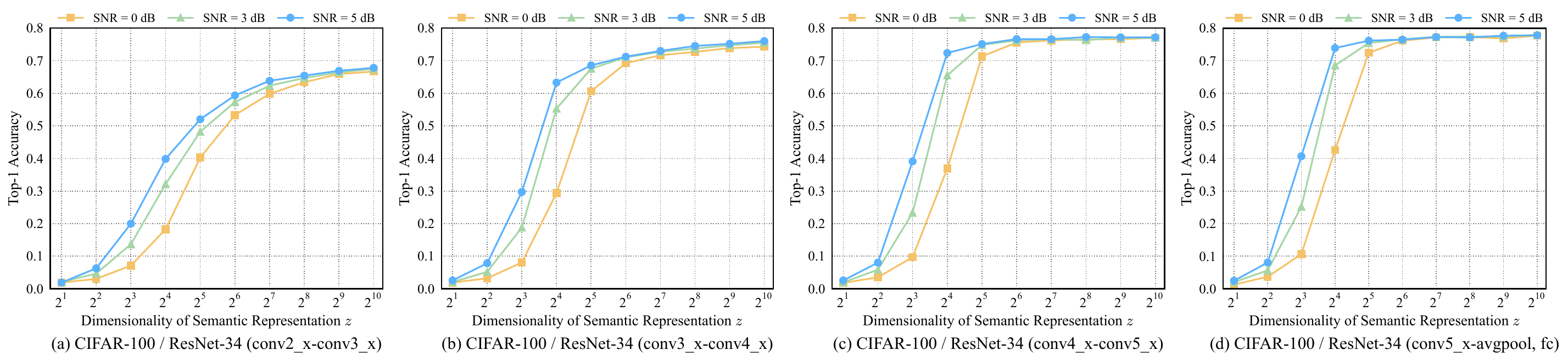}
	\end{center}
    \vspace{-5mm}
	\caption{Comparison of Top-1 Accuracy on CIFAR-100 using ResNet-34 across different $z$ dimensions, SNR levels, and split points. The results for SNR = 5 dB and $N_c=1024$ are consistent with those in Table \ref{flops}.}
        \label{fig:cifar100}
\end{figure*}

\begin{table*}[htbp]
\caption{Comparison of FLOPs, Parameters, and Top-1 Accuracy on CIFAR-100 using ResNet-34 at different split points. Gray shading denotes the same split point in Fig.~\ref{fig:resnet}. Both FLOPs and Parameters are measured in M (million).}
\vspace{-2mm}
\label{flops}
\centering
\begin{tabular}{c|cccc|cccc|c}
\toprule
\multirow{2.5}{*}{\textbf{Split Point}} & \multicolumn{4}{c|}{\textbf{Transmitter (Edge Device)}} & \multicolumn{4}{c}{\textbf{Receiver (Cloud Device)}} & \multicolumn{1}{|c}{\multirow{2.5}{*}{\textbf{Top-1 Accuracy}}} \\ \cmidrule(lr){2-3} \cmidrule(lr){4-5} \cmidrule(lr){6-7} \cmidrule(lr){8-9} & \textbf{FLOPs} & \textbf{Prop. (\%)} & \textbf{Param.} & \textbf{Prop. (\%)} & \textbf{FLOPs} & \textbf{Prop. (\%)} & \textbf{Param.} & \textbf{Prop. (\%)} \\ \midrule\midrule
SP-1 (conv1-conv2\_x) & 2.82 & 0.12 & 0.067 & 0.25 & 2,298.53 & 99.88 & 26.55 & 99.75 & 0.4290 \\ \rowcolor{gray!30}
SP-2 (conv2\_x-conv3\_x) & 229.31 & 9.96 & 0.29 & 1.08 & 2,072.04 & 90.04 & 26.33 & 98.92 & 0.6781 \\
SP-3 (conv3\_x-conv4\_x) & 515.57 & 37.83 & 1.47 & 5.61 & 847.30 & 62.17 & 24.69 & 94.39 & 0.7602 \\
SP-4 (conv4\_x-conv5\_x) & 953.88 & 79.12 & 8.41 & 35.35 & 251.71 & 20.88 & 15.39 & 64.65 & 0.7715 \\
SP-5 (conv5\_x-avgpool, fc) & 1,167.79 & 98.93 & 21.78 & 93.06 & 12.63 & 1.07 & 1.62 & 6.94 & \textbf{0.7781} \\ \bottomrule
\end{tabular}
\end{table*}

We remark that different split points correspond to different practical scenarios. For example, SP-0 is equivalent to directly transmitting the raw image (i.e., receiver-side inference), while SP-6 corresponds to transmitting only the label (i.e., transmitter-side inference). Therefore, we focus on split points from SP-1 to SP-5, which represent task-oriented communication systems with split inference. Since the accuracy for SP-1 is low (0.4298), when edge devices have limited processing power, it is preferable to split at SP-2, which achieves 0.6838 accuracy compared to the baseline 0.7789. If edge devices are more capable, moving the split point to SP-3 can further improve accuracy. Among all split points, SP-4 achieves the highest accuracy of 0.7753, corresponding to 99.5\% of the base model’s performance.

Since device capabilities are a critical consideration in practical networks, we calculate the FLOPs and parameter counts for each module. In the experiments reported in Fig.~\ref{fig:cifar10} and Table~\ref{vanilla}, which use a single fractionally-strided convolution module, we observe that the decompression module at SP-2 is extremely costly, since its parameters and FLOPs increase by approximately 12.8 and 60 times, respectively. To address this, we adopt two fractionally-strided convolution (ConvTranspose2d) modules (as shown in Fig.~\ref{fig:resnet}), to achieve a more practical and balanced FLOP distribution between edge and cloud devices. The updated experimental results are presented in Fig.~\ref{fig:cifar100}, covering four representative split locations after each residual layer block of ResNet-34 (SP-2 to SP-5). As expected, accuracy increases progressively from SP-2 to SP-5. Detailed numerical results on FLOPs, parameter distribution, and accuracy are summarized in Table~\ref{flops}. At SP-2, the transmitter accounts for only about 10\% of total computation. By SP-4, the transmitter’s share rises to 79.12\% of FLOPs and 35.53\% of parameters. At SP-5, nearly all computation (98.93\%) is performed on the transmitter, leaving the receiver’s workload negligible. This evaluation helps identify an optimal split point that satisfies real-world hardware constraints while balancing edge-side efficiency and task accuracy, which follows in the next subsection.

\subsection{Computation and Communication Cost}

Table~\ref{flops} reports the classification accuracy across all model split points. Among them, the earliest split point (SP-1) yields poor performance (less than 0.43), even when using a large semantic feature size ($N_c = 1024$) and under good channel conditions. While increasing $N_c$ could improve accuracy, it also raises the communication burden. When $N_c$ approaches the original input size (e.g., total image pixels), the benefits of semantic compression become negligible. At the other extreme, the final split point (SP-5) offloads only about 1\% of the total computation from the transmitter to the receiver. If the transmitter has sufficient computing capacity, deploying the full model locally is more efficient. In this case, only the classification result (e.g., a label) needs to be sent, minimizing communication cost and avoiding partition overhead. Therefore, we focus on three representative model split points for designing efficient task-oriented communication systems:
\begin{itemize}
    \item \textit{Early split (SP-2):} Offloads approximately 90\% of the model's total FLOPs to the receiver.
    \item \textit{Mid split (SP-3):} Balances computation between the transmitter (roughly 40\%) and the receiver ($\sim$60\%).
    \item \textit{Late split (SP-4):} Allocates approximately 80\% of the computational load to the transmitter.
\end{itemize}

\paragraph{Computation Cost}
We evaluate the normalized computation time of these partitioning strategies. Let $F_M$ be the total FLOPs of the original ResNet-34 model (without splitting). Note that $F_{M_t} + F_{M_r}$ may exceed $F_M$ due to the additional layers inserted during model partitioning. The normalized computation time relative to running the full model at the transmitter is defined as:
\begin{equation}
    \text{Normalized } T_{\text{comp}} = \frac{F_{M_t}}{F_M} + \beta \cdot \frac{F_{M_r}}{F_M},
\end{equation}
where $\beta = \frac{\alpha_r}{\alpha_t}$ is the ratio of per-FLOP processing time between the receiver and transmitter.

\begin{figure}[htbp]
	\begin{center}
	\includegraphics[width=0.5\linewidth]{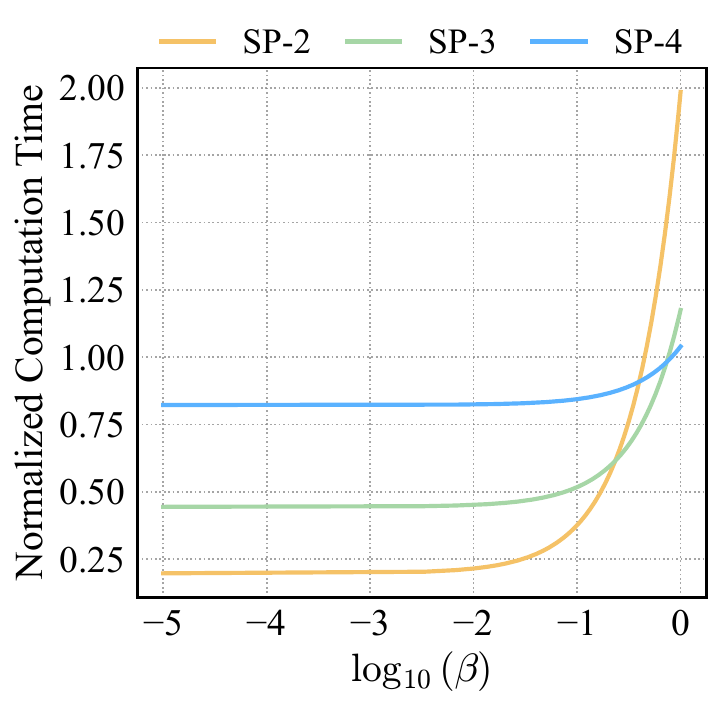}
	\end{center}
    \vspace{-6mm}
	\caption{Normalized $T_{\mathrm{comp}}$ as a function of $\log_{10}(\beta)$.}
    \label{fig:computation_cost}
\end{figure}

In practice, $\beta$ depends on the hardware. Edge devices (e.g., microcontrollers or mobile CPUs) are generally much slower than cloud servers or GPUs. FLOP latency can range from microseconds on edge hardware to picoseconds on cloud platforms, corresponding to a 4–5 order-of-magnitude difference. Therefore, we evaluate $T_{\text{comp}}$ for $\beta$ values ranging from $10^{-5}$ to 1. Figure~\ref{fig:computation_cost} plots the normalized computation cost against $\log_{10}(\beta)$ for the selected split points.

As expected, when $\beta$ is small (i.e., the receiver is significantly faster), offloading computation via partitioning greatly reduces overall computation time. For instance, when $\beta < 10^{-3}$, the Early, Mid, and Late splits achieve computation cost reductions of approximately 80\%, 56\%, and 18\%, respectively. However, as $\beta$ increases and approaches 1, the transmitter and receiver offer similar speed, and the benefit of offloading diminishes. In such cases, partitioning overhead may even lead to increased total computation cost.

\paragraph{Communication Cost}

\begin{figure}[htbp]
	\begin{center}
	\includegraphics[width=0.5\linewidth]{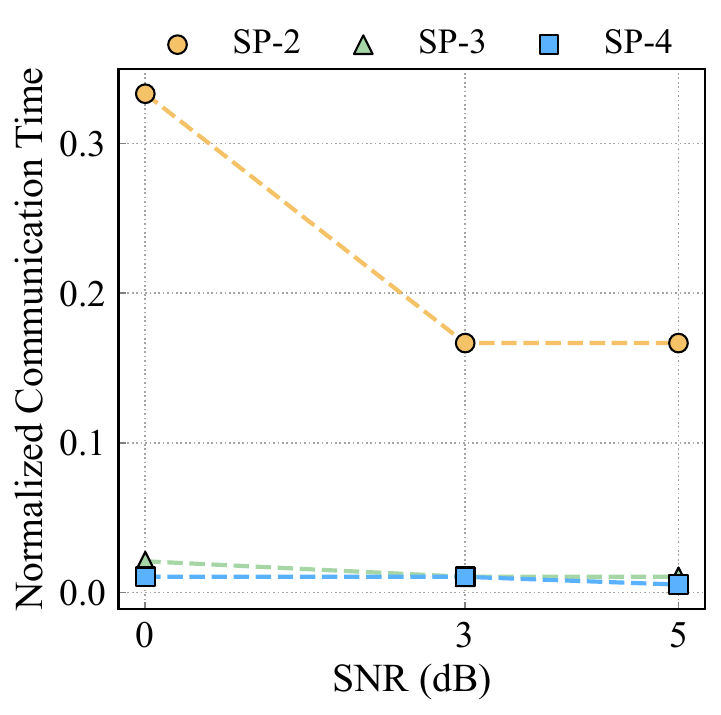}
	\end{center}
    \vspace{-6mm}
	\caption{Normalized communication time under different SNR levels.}
    \label{fig:normalized_communication_cost}
\end{figure}

We evaluate communication efficiency by measuring the transmission time of the semantic feature vector $z$, normalized by the time required to send the original images. Figure~\ref{fig:normalized_communication_cost} shows the normalized communication time for features of dimension $N_c$, where $N_c$ is the minimum size needed to achieve a Top-1 Accuracy of 0.66 (approximately 85\% of the baseline ResNet-34 accuracy without model splitting).

The results show that all split configurations could significantly reduce communication cost compared with transmitting raw input data. However, the Early split (SP-2) requires higher overhead than Mid and Late splits to reach the same accuracy level. This reveals a core trade-off: early partitioning reduces computation at the transmitter but increases communication load. Furthermore, as channel conditions degrade, larger semantic features are needed to preserve accuracy. This highlights the importance of dynamically adapting the dimension of semantic feature, particularly in real-world deployments in changing wireless conditions.

\section{Conclusion and Outlook}

This work demonstrates that under the AWGN channel and varying dimensionalities, task-oriented communication implemented through deep learning model partitioning and semantic compression can achieve a highly efficient balance between classification accuracy, computational load, and communication cost. Experimental results with elaborative analysis provide practical design insights for real-time inference systems in bandwidth- and energy-constrained environments. Future research will focus on developing advanced feature compression and evaluation that dynamically adjusts model partitions and semantic feature dimensions in more realistic wireless scenarios. We also plan to extend this framework to more complex tasks, such as object detection and multi-modal reasoning, and to incorporate energy-aware and privacy-preserving mechanisms for real-world deployment.

\bibliographystyle{IEEEtran}
\bibliography{main}

@book{wcom,
author = {Tse, David and Viswanath, Pramod},
title = {Fundamentals of wireless communication},
year = {2005},
isbn = {0521845270},
publisher = {Cambridge University Press},
address = {USA}
}

@ARTICLE{6773024,
  author={Shannon, C. E.},
  journal={The Bell System Technical Journal}, 
  title={A mathematical theory of communication}, 
  year={1948},
  volume={27},
  number={3},
  pages={379-423}}

@ARTICLE{9919752,
  author={Uysal, Elif and others},
  journal={IEEE Network}, 
  title={Semantic Communications in Networked Systems: A Data Significance Perspective}, 
  year={2022},
  volume={36},
  number={4},
  pages={233-240}}

@ARTICLE{9955525,
  author={Gündüz, Deniz and others},
  journal={IEEE Journal on Selected Areas in Communications}, 
  title={Beyond Transmitting Bits: Context, Semantics, and Task-Oriented Communications}, 
  year={2023},
  volume={41},
  number={1},
  pages={5-41}}

@ARTICLE{10855638,
  author={Getu, Tilahun M. and Kaddoum, Georges and Bennis, Mehdi},
  journal={Proceedings of the IEEE}, 
  title={Semantic Communication: A Survey on Research Landscape, Challenges, and Future Directions}, 
  year={2024},
  volume={112},
  number={11},
  pages={1649-1685}}

@ARTICLE{10554663,
  author={Chaccour, Christina and Saad, Walid and Debbah, Mérouane and Han, Zhu and Vincent Poor, H.},
  journal={IEEE Communications Surveys \& Tutorials}, 
  title={Less Data, More Knowledge: Building Next-Generation Semantic Communication Networks}, 
  year={2025},
  volume={27},
  number={1},
  pages={37-76}}

@inproceedings{10.1145/3037697.3037698,
author = {Kang, Yiping and others},
title = {Neurosurgeon: Collaborative Intelligence Between the Cloud and Mobile Edge},
year = {2017},
isbn = {9781450344654},
publisher = {Association for Computing Machinery},
address = {New York, NY, USA},
booktitle = {Proceedings of the Twenty-Second International Conference on Architectural Support for Programming Languages and Operating Systems},
pages = {615–629},
numpages = {15},
location = {Xi'an, China},
series = {ASPLOS '17}
}

@INPROCEEDINGS{9145068,
  author={Shao, Jiawei and Zhang, Jun},
  booktitle={2020 IEEE International Conference on Communications Workshops (ICC Workshops)}, 
  title={BottleNet++: An End-to-End Approach for Feature Compression in Device-Edge Co-Inference Systems}, 
  year={2020},
  volume={},
  number={},
  pages={1-6}}

@INPROCEEDINGS{10226176,
  author={Sagduyu, Yalin E. and Ulukus, Sennur and Yener, Aylin},
  booktitle={IEEE Conference on Computer Communications Workshops (INFOCOM WKSHPS)}, 
  title={Age of Information in Deep Learning-Driven Task-Oriented Communications}, 
  year={2023},
  volume={},
  number={},
  pages={1-6}}

@INPROCEEDINGS{7780459,
  author={He, Kaiming and Zhang, Xiangyu and Ren, Shaoqing and Sun, Jian},
  booktitle={2016 IEEE Conference on Computer Vision and Pattern Recognition (CVPR)}, 
  title={Deep Residual Learning for Image Recognition}, 
  year={2016},
  volume={},
  number={},
  pages={770-778}}

@techreport{krizhevsky2009learning,
  title        = {Learning Multiple Layers of Features from Tiny Images},
  author       = {Alex Krizhevsky},
  institution  = {University of Toronto},
  year         = {2009},
  url          = {},
  note         = {Technical Report}
}

@inproceedings{pytorch,
  author = {Paszke, Adam and others},
  booktitle = {NIPS 2017 Workshop on Autodiff},
  location = {Long Beach, California, USA},
  title = {Automatic Differentiation in PyTorch},
  url = {},
  year = 2017
}

@ARTICLE{10287247,
  author={Sun, Yaping and Chen, Hao and Xu, Xiaodong and Zhang, Ping and Cui, Shuguang},
  journal={IEEE Transactions on Wireless Communications}, 
  title={Semantic Knowledge Base-Enabled Zero-Shot Multi-Level Feature Transmission Optimization}, 
  year={2024},
  volume={23},
  number={5},
  pages={4904-4917},
  doi={10.1109/TWC.2023.3323380}}

\end{document}